\documentstyle[11pt,newpasp,twoside,epsf]{article}
\def\edcomment#1{\iffalse\marginpar{\raggedright\sl#1\/}\else\relax\fi}
\reversemarginpar

\newcommand{\etal}{{\sl et~al.}}

\newcommand{\fuse}{{\sl FUSE}}
\newcommand{\hs}{{\sl HST}}
\newcommand{\hu}{{\sl HUT}}
\newcommand{\orf}{{\sl ORFEUS}}

\newcommand{\teff}{\mbox{$T_{\rm eff}\:$}}

\begin{document}

\title
{\fuse \ Observations of the Hottest DA White Dwarfs}

\author{M.A. Barstow, M.R. Burleigh, N.P. Bannister}  
\affil{Department of Physics and Astronomy, University of Leicester,
University Road, Leicester LE1 7RH, UK}
\author{J.B. Holberg}
\affil{Lunar and Planetary Laboratory, University of Arizona, Tucson, 
AZ 85721, USA}
\author{I. Hubeny}
\affil{Laboratory for Astronomy and Solar Physics, NASA/GSFC, Greenbelt,
Maryland, MD 20711 USA}
%\author{R. Napiwotzki}
%\affil{Dr. Remesi-Strenwarte, Sternwartstr. 7, D-96049 Bamberg, Germany}

\begin{abstract}

We report early \fuse \ observations of the very hot DA white dwarfs
PG1342+444 and REJ0558$-$371. Detection of photospheric absorption lines
allows us to estimate the abundances of C, O, Si, P, S and Fe, the first
measurements reported for DA stars at such high temperatures.
Values of \teff \ and log g determined for PG1342+444 
from the Lyman line series disagree with the results 
of the standard Balmer line analysis, 
an issue that requires further investigation.

\end{abstract}

\section{Introduction}

Very few true white dwarfs are known to have effective temperatures above
$\approx 70000$K and, therefore, the proposed direct evolutionary link
between H-rich CSPN and white dwarfs has hardly been explored. 
The discovery of
several very hot DA white dwarfs with temperatures in excess of 70000K
(e.g. REJ1738+665 - 90000K, Barstow \etal \ 1994; EGB1 - 100000K,
Napiwotzki \& \ Sch\"onberner 1993; WDHS1 - 100000 to 160000K, Liebert,
Bergeron \& \ Tweedy 1994), through various observational programmes,
provides an opportunity to study this upper range of the DA cooling
sequence. 

While Balmer line analyses (including the effect of metals) provide
important measurements of the DA temperature scale, there remains a
serious problem in their use to study the hottest DA stars. At effective
temperatures in excess of $\approx 50000$K, the photospheric hydrogen is
almost completely ionized. Hence, the populations of the levels involved
in the Balmer line transitions are severely depleted and the observed
line strengths extremely weak. As a result, at high temperatures, the
values of \teff \ become increasingly poorly determined. The use of the
Lyman series lines has provided an additional tool to determine \teff \
and log g but only a limited number of observations have been available
from the \orf \ and Hopkins Ultraviolet Telescope (\hu ) missions (e.g.
Dupuis \etal \ 1998; Finley \etal \ 1997). 

The availability of the Far Ultraviolet
Spectroscopic Explorer (\fuse ),
covering the $\approx
900-1200$\AA \ wavelength range, presents us with an opportunity of
obtaining many more Lyman series spectra for white dwarfs. 
We present \fuse \ observations of the very hot DA white dwarfs PG1342+444
and REJ0558$-$376, comparing the Balmer and Lyman series
measurements of \teff . An important result is the first
detection of photospheric  OVI lines in the
atmosphere of a DA white dwarf.

\section{Far UV spectroscopy with \fuse }

\subsection{Observations and data reduction}

The \fuse \ mission was placed in low Earth orbit on 1999 June 24. After
several months of in-orbit checkout and calibration activities, science
observations began during 1999 December. An overview of the  \fuse \ mission 
has been given by Moos \etal \ (2000) and the spectrograph performance
is described in detail by Sahnow \etal \  (2000). Further
useful information is included in the \fuse \ Observer's Guide (Oegerle
\etal \ 1998) which can be found on the \fuse \ website
(http://fuse.pha.jhu.edu) along with other technical documentation.

Several problems have been encountered during early in-orbit operations of the
\fuse \ satellite which need to be taken account of in the data reduction
process. Maintaining the coalignment of the individual spectral channels
has been difficult, probably due to thermal effects. Sometimes, a target
may completely miss the aperture of one or more channels, while being
well-centred in the others. In addition, even if the channels are
well-aligned at the beginning of an observation, the target may
subquently drift out of any of the apertures. There are also several
effects associated with the detectors and electronics, including
deadspots and fixed-pattern efficiency variations. Potentially, these can
lead to spurious absorption features in the processed data if they cut
the dispersed spectrum at any point.

The FUSE observations of PG~1342+444 and RE~J0558$-$376 were obtained
during 2000 January 11/12 and 1999 December 10 respectively. Inspection
of the combined \fuse \ spectra of each star shows the expected flux
levels but reveals some discontinuities in the flux matching particular
channel wavelength ranges. This indicates that in some channels the
actual exposure time achieved does not match the nominal value, possibly
arising from the source drifting out of the science aperture.
Consequently, we examined each individual extracted and calibrated
spectrum and developed our own approach to combining the various
exposures and channels based on an assessment of the data quality. Since,
the nominal wavelength binning ($\approx 0.006$\AA) oversamples the true
resolution by a factor of 2-3, all the spectra were re-binned to 0.02\AA
\ pixels for analysis. Barstow \etal \ (2000, in preparation) 
describe our procedure in
much more detail. 

Figure~1 shows the final merged spectra
constructed from the best exposures in each spectral channel. A broad dip
in the region above $\approx 1160$\AA \ can be seen in each case, caused
by a region of low instrumental efficiency known as the ``worm'' and
which is not currently taken into account in the flux calibration. The
short wavelength region (SiC channel) of the RE~J0558$-$276 spectrum,
below 990\AA , is considerably noisier than the longer wavelength (LiF
channel) data due to periods
when the source drifted out of the SiC aperture during the exposure or
was missed completely during acquisition of the target. A number of
apparent emission features seen in PG~1342+444 are not stellar features
but arise from geocoronal emission.

\begin{figure}
\plottwo{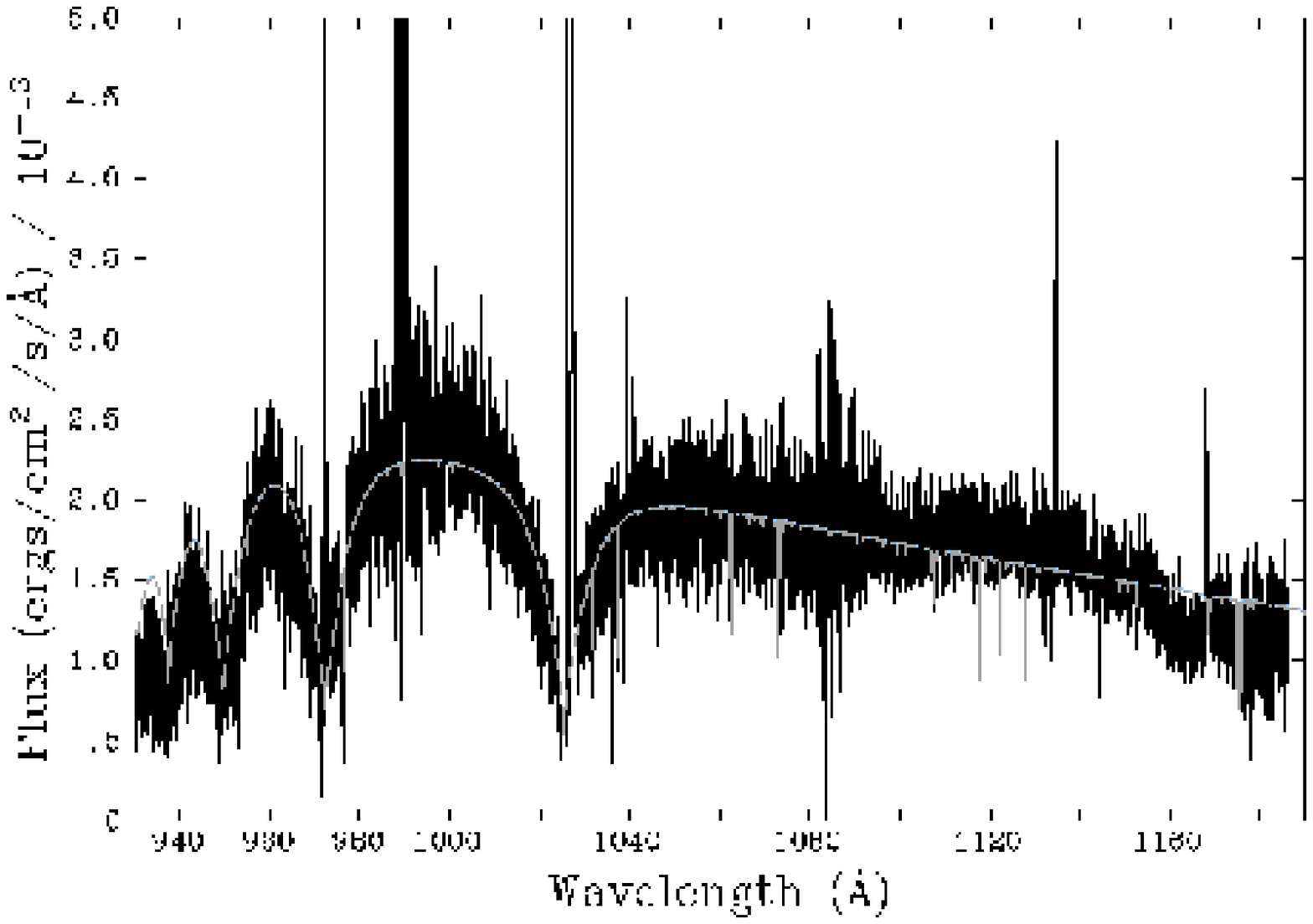}{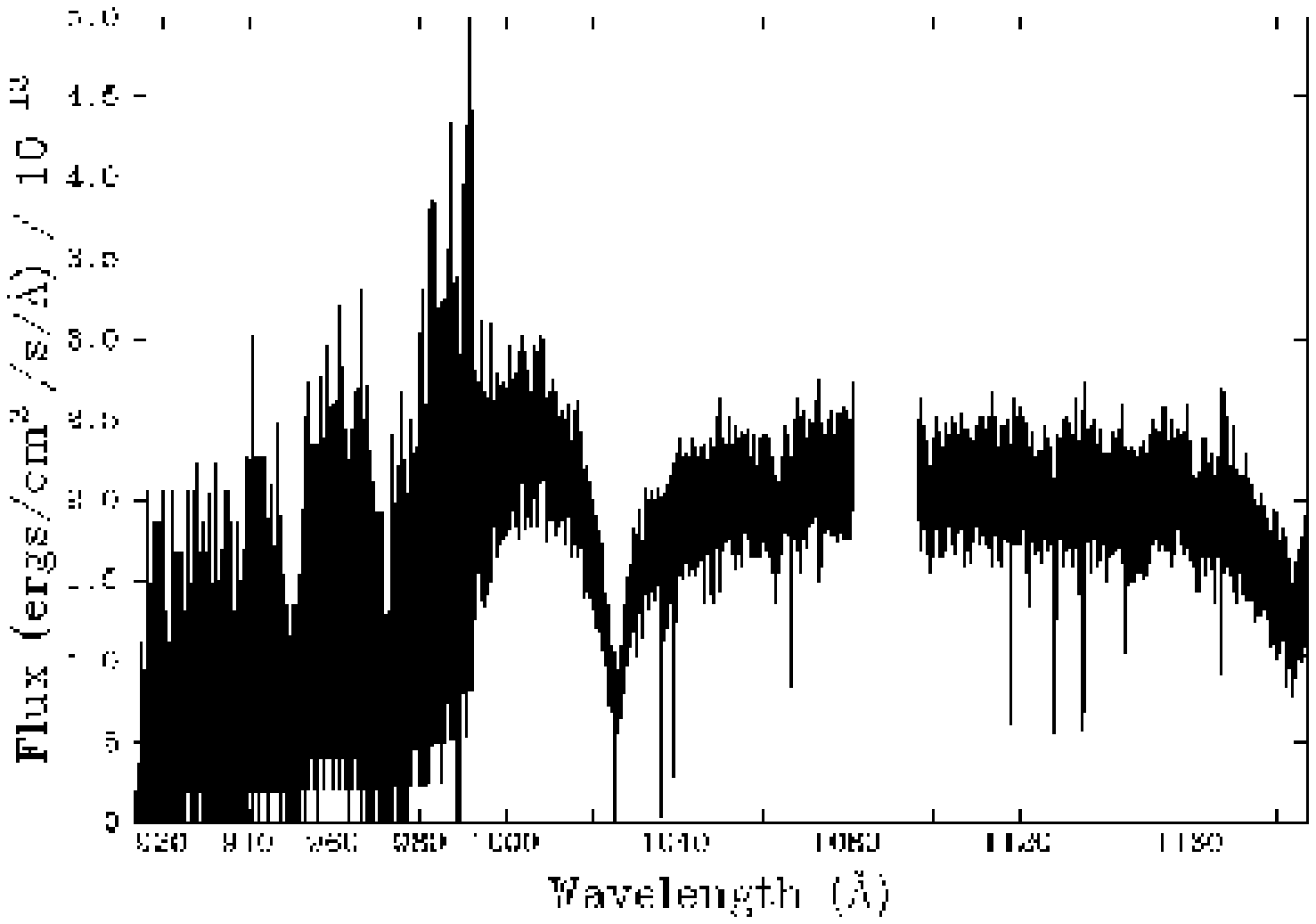}
\caption{Composite \fuse \ spectra of PG~1342+444
(left) and RE~J0558$-$371 (right). A number of features can be see
including the H Lyman line series and some strong photospheric in
interstellar lines of heavier elements. }
\end{figure}

\subsection{Detection of 
photospheric and interstellar features in \fuse \ spectra}

It is not surprising that absorption lines from both the local ISM and
the photosphere of the stellar sources are seen in the \fuse \ spectra.
Such features have already been reported in the \orf \ spectra of
G191-B2B and RE~J0457$-$281 (Vennes \etal \ 1996), including the
detection of S and P, which are not seen in any other wavebands. However,
\orf \ was only able to observed a handful of WDs in its limited duration
mission. With a planned mission lifetime of at least 2 years and a
superior spectral resolution ($\lambda/\delta \lambda=20,000\ {\rm cf}\ 5,000$),
\fuse \ will provide a more comprehensive study of hot WDs and is capable
of detecting weaker features.

The compressed wavelength scale of Figure~1 renders invisible
all but the strongest absorption features. However, closer examination
reveals many absorption lines. For example, the 1120 to 1130 region of
the PG~1342+444 spectrum shows photospheric SiIV and PV together with
interstellar CI and FeII (figure~2). These same species are
seen in RE~J0558$-$371, while photospheric CIV, SVI and FeVI are also
detected in both objects.

Possibly the most interesting result is the presence of the OVI resonance
absorption lines at 1031.912\AA \ and 1037.613\AA \ in both stars 
(figure~2), the first detection of OVI in any
H-rich WD. This results also represents the first observation of these
resonance features in any WD. Such a high oxygen ionization state is
typically associated with the very hot PG1159 stars, being detected at
optical wavelengths in those objects (e.g. Werner, Heber, \& \ Hunger
1991). Hence, its appearance in a DA spectrum might appear to be
something of a surprise since, at \teff ~70,000K, PG~1342+444 and
RE~J0558$-$371 are considerably cooler. Nevertheless, the presence of OVI
features is predicted by the model atmosphere calculations.
Interestingly, OVI was not detected in the \orf \ spectra of the cooler
objects G191-B2B and RE~J0457$-$281. Table~1 lists the
photospheric abundances of all elements detected in the \fuse \ spectral
range, determined using the latest non-LTE model calculations, and
compares these with values measured from \orf \ data for G191-B2B.

\begin{figure}
\plottwo{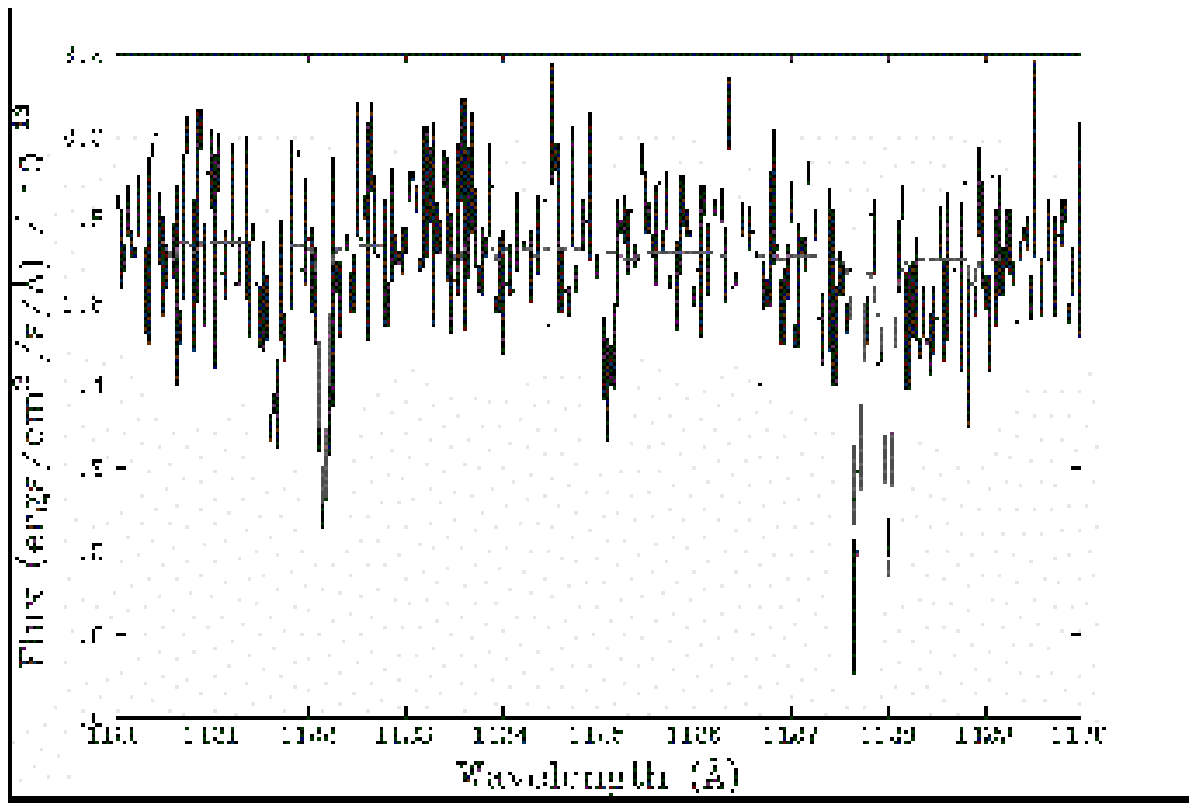}{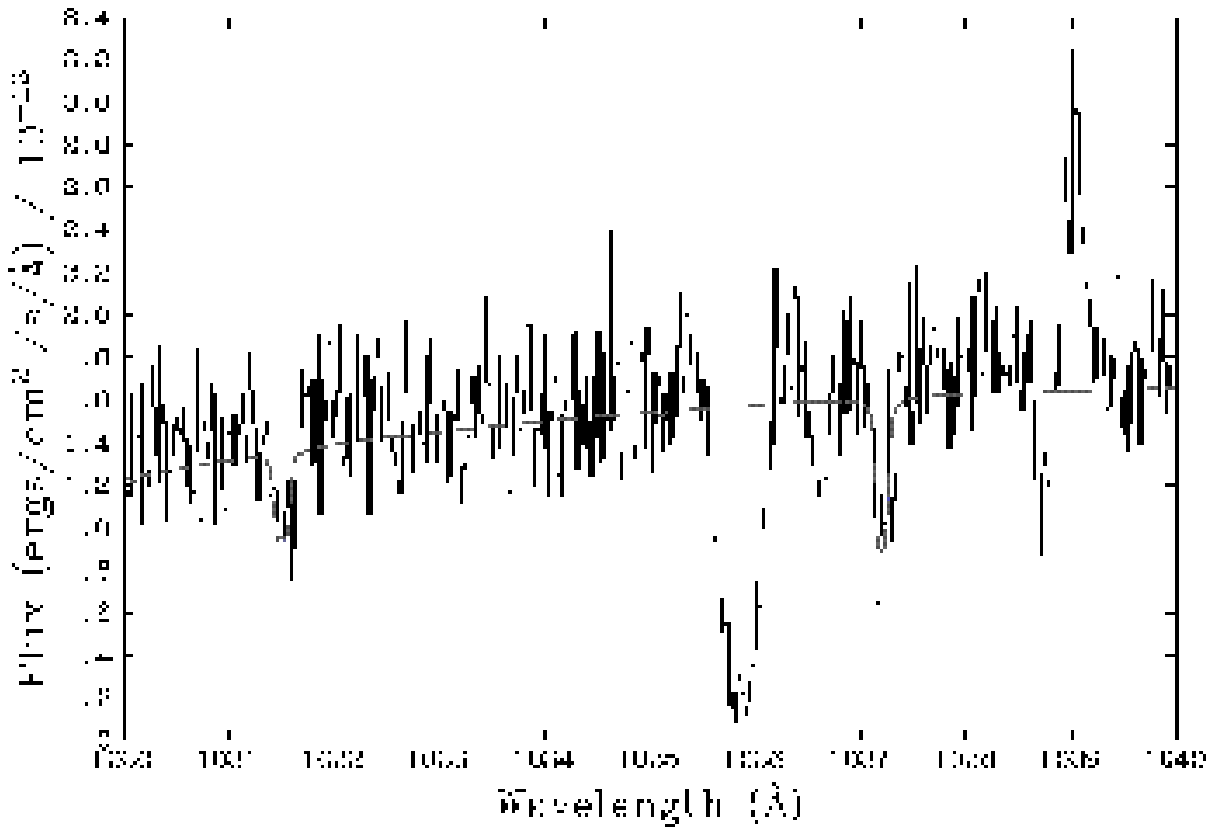}
\caption{Left: A 10\AA \
region of the \fuse \ spectrum of PG~1342+444, showing photospheric SiIV
(1122.485/1128,340\AA ) and PV (1128.010\AA ) features together with 
interstellar CI (1128.079\AA ) and FeII (1125.4478\AA ). The data are
represented by the histogram while the smooth curve is a synthetic
spectrum which best matches the strength of the photospheric lines.
Right: Detection of photospheric OVI (1031.912/1037.613\AA ) 
in the \fuse \ spectrum of PG~1342+444.
Also visible are strong CII (1036.3367/1037.0179\AA ) interstellar features. 
}
\end{figure}

\begin{table}
\caption{Heavy element abundances (relative to hydrogen) 
determine from \fuse \ observations of PG~1342+444 and RE~J0558$-$371,
compared to our measurements of the G191-B2B \orf \ and \hs \ STIS spectra.}
\begin{tabular}{lccc}
Element & &Abundance (N$_{\rm elem}$/H) \\
       & PG~1342+444 & RE~J0558$-$371 & G191-B2B\\
Carbon & $1.0\times 10^{-5}$ & $8.0\times1 0^{-5}$ & $4.0\times 10^{-7}$ \\
Oxygen & $2.0\times 10^{-6}$ & $3.0\times 10^{-6}$ & $1.0\times 10^{-6}$\\
Silicon & $1.0\times 10^{-6}$ & $2.0\times 10^{-6}$ & $3.0\times 10^{-7}$\\
Phosphorus & $1.0\times 10^{-8}$& $2.0\times 10^{-8}$ & $2.5\times 10^{-8}$\\
Sulphur & $2.0\times 10^{-6}$ & $2.5\times 10^{-7}$ & $3.2\times 10^{-7}$\\
Iron & $1.0\times 10^{-5}$ & $1.0\times 10^{-5}$ & $1.0\times 10^{-5}$\\
\end{tabular}
\end{table}

%\begin{figure}
%\plotone{}
%\caption{
%}
%\label{}
%\end{figure}

\section{Determination of Temperature and Gravity}

To measure the values of \teff \ and log g from the Lyman lines visible
in the \fuse \ spectra, we have adapted the standard technique we have
used previously for Balmer line studies (e.g. Barstow \etal \ 1997b).
All the lines ($\beta, \gamma, \delta, \epsilon $) included are fit 
simultaneously 
and an independent normalisation constant was applied to each, ensuring that 
the result was independent of the local slope of the continuum and reducing 
the effect of any systematic errors in the flux calibration of the spectra.  
We summarise the results of the Lyman line analyses in Table~2 and
compare these with new Balmer line studies, utilising the same non-LTE
heavy element-rich  models, and earlier Balmer line work based on pure H
atmospheres.

\begin{table}
\caption{Effective temperature and surface gravity measured for
PG~1342+444 and REJ~0558$-$371 from Lyman and Balmer lines.
$1\sigma $ uncertainties are in brackets.}
\begin{tabular}{lcccc}
     & PG~1342+444 && RE~J0558$-$371 \\
     & \teff (K) & log g & \teff (K) & log g \\
Nominal (pure H) & 79000 & 7.82 & 70000 & 7.37\\
Lyman & 55800 (660) & 8.00 (0.02) & 61100 (660) & 7.61 (0.05)\\
Balmer& 66750 (2500)  & 7.93 (0.01) & 60800 (2500) & 7.55 (0.15)\\
\end{tabular}
\end{table}

\section{Discussion}

As has been reported for cooler hot DA WDs (Barstow, Hubeny and Holberg
1998), the values of \teff \ measured with fully line blanketed,
heavy element-rich, non-LTE model atmospheres are significantly lower than
those reported for pure H analyses. Hence, we must conclude that,
overall, the temperature scale of the very hottest white dwarfs such as 
PG~1342+444 and RE~J0558$-$371 is lower than currently believed. Since
these stars, together with other objects in our \fuse \ programme, were
selected on the basis of their location in the temperature range between
the hot DAs, such as G191-B2B, and the CPN, this result seems to reopen a
gap between the CPN and the white dwarf cooling sequence. However, no CPN
have yet been subjected to the same kind of analysis as that applied to these 
hot DAs. It seems likely that a similar reduction in \teff \ will be
found when such an analysis is carried out.

It is interesting to note that, the uncertainties in the values of \teff \ 
determined from the Lyman line work are lower than for the Balmer lines.
This probably arises from the fact that the Balmer lines become
increasingly weak at these high temperatures, whereas the Lyman lines are
much less affected by this problem. The respective Lyman and Balmer line
measurements are in good agreement for RE~J0558$-$371 but there is a
10000K inconsistency between the two PG~1342+444 results. At the moment,
this is difficult to explain as there are few examples of similar analyses
with which to compare these results. The relative behaviour of Lyman and Balmer
lines can be influenced by assumptions built into the model atmosphere
calculations, although in this case we might expect to see differences
for all stars. Alternatively, there may be signficant interstellar H
opacity in PG~1342+444 which has not been properly accounted for in the
analysis. Such material would increase the strength of the Lyman lines,
leading to a lower value of \teff .

\section{Conclusion}

We have presented observations of some of the hottest known white dwarfs
made with the \fuse \ spectrometer. Detection of photospheric CIV, SiIV,
OVI, PV, SVI and FeVI absorption lines allows us to estimate the
abundances of these elements, the first such measurements reported for DA
WDs at these high temperatures. The observed abundances are similar to
those found in the somewhat cooler star G191-B2B (\teff $\approx
54000$K). However, PG~1342+444 seems to have a large excess of C when
compared to both REJ~0558$-$371 and G191-B2B. The OVI resonance lines are
uniquely present in the \fuse \ wavelength range, and their detection in
the PG~1342+444 and RE~J0558$-$371 spectra is the first such report for
any hot white dwarf.

The Balmer and Lyman line studies yield values of \teff \ lower than
reported for pure H Balmer line studies, as might be expected. The
inconsistency between the PG~1342+444 Lyman and Balmer results sn an
issue that requires further investigation. Forthcoming observations of
the remaining targets in our sample will help examine whether or not the
discrepancy originates in the model atmosphere calculations or arises
from systematic effects due to interstellar absorption, or in the
\fuse \ spectrometer calibration.

\section*{Acknowledgements}

The work reported in the papers was based on observations
made with the \fuse \ observatory, operated on behalf of NASA
by Johns Hopkins University. MAB, MRB and NPB were supported by PPARC, UK.


\begin{references}
 
\reference Barstow, M.A. \etal , 1994, MNRAS, 271, 175

\reference Barstow, M.A., Holberg, J.B., Cruise, A.M., \& \ Penny A.J., 
1997, MNRAS, 290, 505

\reference Barstow, M.A., Hubeny, I., and Holberg, J.B., 1998, MNRAS,
299, 520.

\reference Dupuis, J., Vennes, S., Chayer, P., Hurwitz, M., and Bowyer, S.,
1998, ApJL, 500, L45.

\reference Finley, D.S., Koester, D., Kruk, J.W., Kimble, R.A., and
Allard, N.F., 1997, in Proc 10th European Workshop on White Dwarfs, eds.
J. Isern, M. Hernanz \& \  E. Garcia-Berro (Dordrecht: Kluwer), 245.

\reference Liebert, J., Bergeron, P., \& \  Tweedy, R.W., 1994, ApJ, 424, 817

\reference Moos, H.W., \etal , 2000, ApJL, 538, 1.

\reference Napiwotzki, R., Sch\" onberner, D., 
1993, In `White Dwarfs: Advances in Observation and Theory', ed. M.A. Barstow, Kluwer, Dordrecht, 99

\reference Oegerle, W.R., \etal , 2000, The Fuse Observers Guide, Version 2.1.

\reference Sahnow, D.J., \etal , 2000, ApJL, 538, 7.

\reference Vennes, S., Chayer, P., Hurwitz, M., \& \ Bowyer, S., 1996, ApJ,
468, 989

\reference Werner, K., Heber, U., and Hunger, K., 1991, A\& A, 244, 437.

\end{references}
\end{document}